\pdfoutput=1

\documentclass[twocolumn,aps,pra]{revtex4}

\usepackage{graphicx}
\usepackage{amsmath}
\usepackage{amssymb}

\begin{document}


\title{Maximum quantum nonlocality between systems that never interacted}


\author{Ad\'an Cabello}
 \affiliation{Departamento de F\'{\i}sica Aplicada II, Universidad de
 Sevilla, E-41012 Sevilla, Spain\\E-mail: adan@us.es\\Phone: 34-954-556671\\Fax: 34-954-556672}


\date{\today}


\begin{abstract}
We show that there is a stronger form of bipartite quantum nonlocality in which systems that never interacted are as nonlocal as allowed by no-signaling. For this purpose, we first show that nonlocal boxes, theoretical objects that violate a bipartite Bell inequality as much as the no-signaling principle allows and which are physically impossible for most scenarios, are feasible if the two parties have 3 measurements with 4 outputs. Then we show that, in this case, entanglement swapping allows us to prepare mixtures of nonlocal boxes using systems that never interacted.
\end{abstract}


\maketitle


\section{Introduction}


The violation of the Clauser-Horne-Shimony-Holt (CHSH) Bell inequality \cite{Bell64,CHSH69} has been experimentally observed (up to loopholes \cite{CS12}) many times \cite{ADR82,TBZG98,WJSWZ98,MMMOM08} and has a number of applications \cite{BZPZ04,BHK05,PAMBMMOHLMM10}. However, although counterintuitive, the maximum quantum violation in this case \cite{Cirelson80} is not as large as the one allowed by the no-signaling principle \cite{PR94}. The aim of this Letter is to show that, under the appropriate conditions, quantum mechanics allows pairs of systems that never interacted to produce correlations as nonlocal as allowed by no-signaling.


\section{Quantum nonlocal boxes}


Nonlocal boxes (NBs) are theoretical objects shared by two parties, Alice and Bob, that allow them to violate a Bell inequality as much as is allowed by the no-signaling principle. NBs were introduced to illustrate that the maximum quantum violation of the CHSH inequality is not as large as allowed by the no-signaling principle \cite{PR94}. For the CHSH and other Bell inequalities, NBs are rendered impossible by some physical principles \cite{VanDam00,PPKSWZ09,OW10,PW12,A12,Cabello12b}.

\begin{figure}[t]
\vspace{-3cm}
 \centerline{\includegraphics[width=0.51\textwidth]{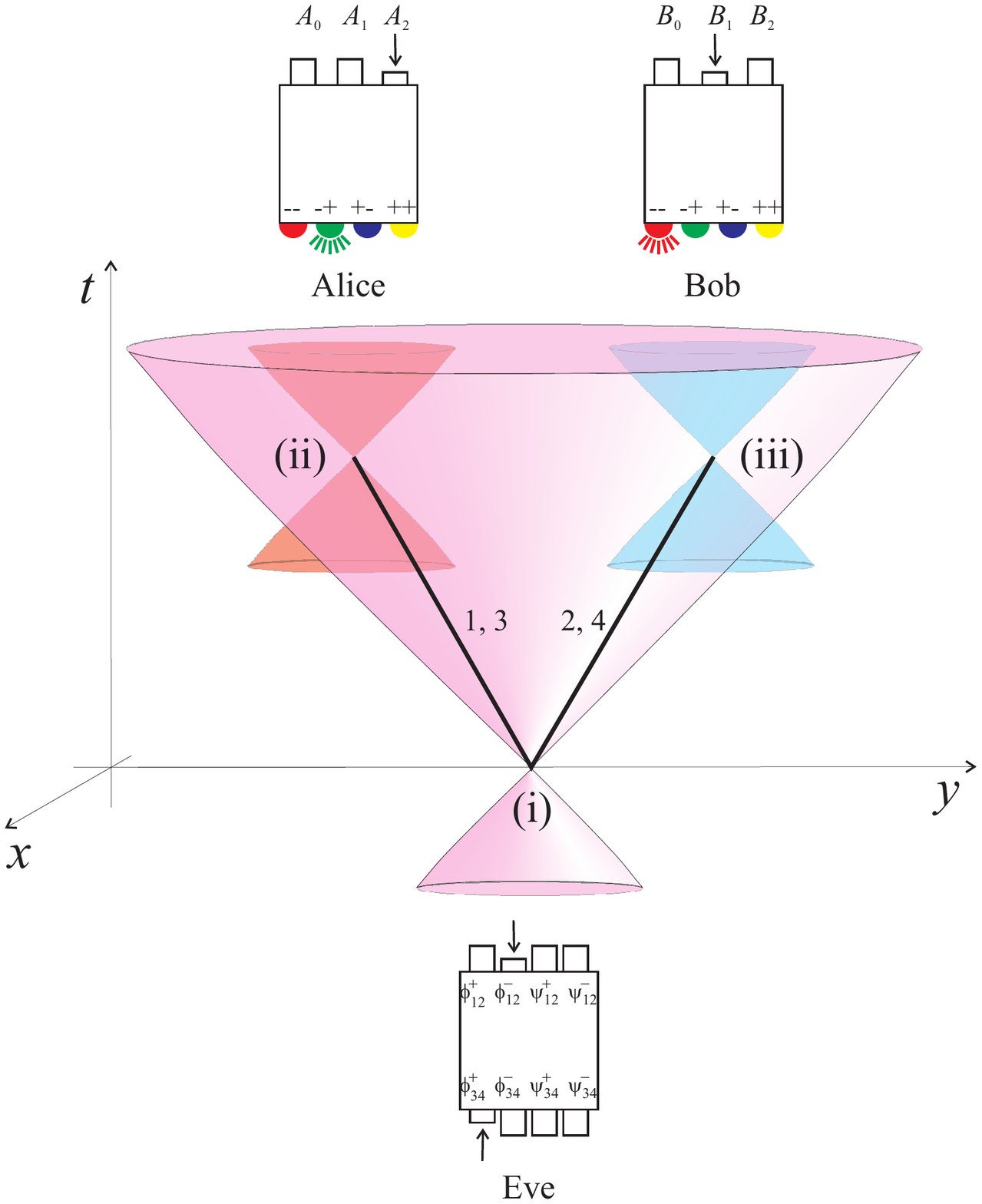}}
\caption{\label{Fig1}{\em Quantum nonlocal boxes.} Space-time diagram of a run of an experiment in which Eve produces a NB. In (i), Eve prepares 4 qubits in a product of Bell states, e.g., qubit 1 and 2 in the state $\phi^-_{12}$ and qubits 3 and 4 in the state $\phi^+_{34}$. In (ii), Alice presses a button, e.g., $A_2$ (leading to observable $A_2$ being measured on qubits 1 and 3) and records the result (e.g., $-+$). In (iii), spacelike separated form (ii), Bob presses a button, e.g., $B_1$ (leading to observable $B_1$ being measured on qubits 2 and 4) and records the result (e.g., $--$). Eve previously set observables $A_i$ and $B_i$ to be those defined in (\ref{Ai}) and (\ref{Bi}), respectively.}
\end{figure}

We first show that, if Alice and Bob can choose between 3 experiments with 4 outcomes, then a third party, Eve, can prepare NBs. We prove it by constructing an explicit example which is easy to follow with the aid of black boxes with buttons; see Fig.~\ref{Fig1}. For each run of the experiment, Alice and Bob randomly choose one out of 3 buttons, representing the experiments they can perform on a system prepared by Eve. Alice's buttons are labeled $A_0$, $A_1$, $A_2$, and Bob's $B_0$, $B_1$, $B_2$. Every time one button is pressed, one light flashes, indicating the corresponding result. There are 4 lights, and we label each of them with two bits: $--$ [meaning $(-1,-1)$], $-+$, $+-$, and $++$. We suppose that there is spacelike separation between the event in which Alice (Bob) presses the button and the event in which Bob (Alice) records the result, so that there cannot be causal influences between them.


Consider the following 16 Bell inequalities:
\begin{equation}
 \label{nonlocalineq}
\beta_i \stackrel{\mbox{\tiny{ LHV}}}{\leq} 7 \stackrel{\mbox{\tiny{ NS}}}{\leq} 9,
\end{equation}
where $i=1,\ldots,16$, $\stackrel{\mbox{\tiny{ LHV}}}{\leq} 7$ indicates that, for any local hidden variable theory, $\beta_i$ is upper bounded by 7, $\stackrel{\mbox{\tiny{ NS}}}{\leq} 9$ indicates that, for any theory satisfying no-signaling, $\beta_i$ is upper bounded by 9, and
\begin{equation}
 \label{beta}
 \begin{split}
 \beta_i=&s_{00} \langle A_0^{10} B_0^{10} \rangle + s_{01} \langle A_0^{01} B_1^{10} \rangle + s_{02} \langle A_0^{11} B_2^{10} \rangle \\
 &+ s_{10} \langle A_1^{10} B_0^{01} \rangle + s_{11} \langle A_1^{01} B_1^{01} \rangle + s_{12} \langle A_1^{11} B_2^{01} \rangle\\
 &+ s_{20} \langle A_2^{10} B_0^{11} \rangle + s_{21} \langle A_2^{01} B_1^{11} \rangle + s_{22} \langle A_2^{11} B_2^{11} \rangle ,
 \end{split}
\end{equation}
where $\langle A_1^{10} B_0^{01} \rangle$ is the mean value of the product of the first bit ($-1$ or $1$) of the result of measuring $A_1$, multiplied by the second bit of the result of $B_0$, and $\langle A_2^{01} B_1^{11} \rangle$ is the mean value of the product of the second bit of the result of $A_2$ times the first bit of the result of $B_1$ times the second bit of the result of $B_1$. For each $\beta_i$, the values of $s_{00},\ldots,s_{22}$ are given in Table \ref{Table2}. It can be checked (e.g., using \texttt{porta} \cite{Porta}), that all these 16 inequalities are tight (i.e., facets of the bipartite 4-measurement, 3-outcome local polytope \cite{Pitowsky89}).


\begin{table}[htb]
\caption{\label{Table2}Coefficients $s_j$ in (\ref{beta}).}
\begin{ruledtabular}
\begin{tabular}{r|rrrrrrrrr}
 & $s_{00}$ & $s_{01}$ & $s_{02}$ & $s_{10}$ & $s_{11}$ & $s_{12}$ & $s_{20}$ & $s_{21}$ & $s_{22}$ \\
\hline
$\beta_{1,16}$ & $\pm 1$ & $\pm 1$ & $1$ & $\pm 1$ & $\pm 1$ & $1$ & $1$ & $1$ & $-1$ \\
$\beta_{2,15}$ & $\pm 1$ & $\pm 1$ & $1$ & $\mp 1$ & $\pm 1$ & $-1$ & $-1$ & $1$ & $1$ \\
$\beta_{3,14}$ & $\pm 1$ & $\mp 1$ & $-1$ & $\pm 1$ & $\pm 1$ & $1$ & $1$ & $-1$ & $1$ \\
$\beta_{4,13}$ & $\pm 1$ & $\mp 1$ & $-1$ & $\mp 1$ & $\pm 1$ & $-1$ & $-1$ & $-1$ & $-1$ \\
$\beta_{5,12}$ & $\pm 1$ & $\pm 1$ & $1$ & $\pm 1$ & $\mp 1$ & $-1$ & $1$ & $-1$ & $1$ \\
$\beta_{6,11}$ & $\pm 1$ & $\pm 1$ & $1$ & $\mp 1$ & $\mp 1$ & $1$ & $-1$ & $-1$ & $-1$ \\
$\beta_{7,10}$ & $\pm 1$ & $\mp 1$ & $-1$ & $\pm 1$ & $\mp 1$ & $-1$ & $1$ & $1$ & $-1$ \\
$\beta_{8,9}$ & $\pm 1$ & $\mp 1$ & $-1$ & $\mp 1$ & $\mp 1$ & $1$ & $-1$ & $1$ & $1$ \end{tabular}
\end{ruledtabular}
\end{table}


Eve sets Alice's $A_0$, $A_1$, and $A_2$ to the following two-qubit observables:
\begin{subequations}
\label{Ai}
\begin{align}
 A_0= & r_{++} |00\rangle \langle00|+ r_{+-} |01\rangle \langle01| \nonumber\\
 &+ r_{-+} |10\rangle \langle10|+ r_{--} |11\rangle \langle11|,\\
 A_1= & r_{++} |++\rangle \langle++|+ r_{+-} |-+\rangle \langle-+| \nonumber\\
 &+ r_{-+} |+-\rangle \langle+-| + r_{--} |--\rangle \langle--|,\\
 A_2= & r_{++} |\chi^+\rangle\langle\chi^+|+ r_{+-} |\chi^-\rangle \langle\chi^-| \nonumber\\
 &+ r_{-+} |\omega^+\rangle\langle\omega^+| + r_{--} |\omega^-\rangle\langle\omega^-|,
\end{align}
\end{subequations}
where
$\sigma_z \left| 0 \right\rangle =
\left| 0 \right\rangle$,
$\sigma_z \left| 1 \right\rangle =
-\left| 1 \right\rangle$,
$\sigma_x \left| \pm \right\rangle = \pm
\left| \pm \right\rangle$, and
\begin{subequations}
\begin{align}
 &|\chi^\pm \rangle = \frac{1}{\sqrt{2}} (|0+\rangle \pm |1- \rangle),\\
 &|\omega^\pm \rangle = \frac{1}{\sqrt{2}} (|1+ \rangle \pm | 0- \rangle)
\end{align}
\end{subequations}
are the common eigenvectors of $\sigma _z\otimes \sigma _x$ and $\sigma _x\otimes \sigma _z$.

Similarly, Eve sets Bob's $B_0$, $B_1$, and $B_2$ to the following two-qubit observables:
\begin{subequations}
\label{Bi}
\begin{align}
 B_0= & r_{++} |0+\rangle \langle0+|+ r_{+-} |0-\rangle \langle0-|\nonumber\\
 &+ r_{-+} |1+\rangle \langle1+|+ r_{--} |1-\rangle \langle1-|,\\
 B_1= & r_{++} |+0\rangle \langle+0|+ r_{+-} |-0\rangle \langle-0| \nonumber\\
 &+ r_{-+} |+1\rangle \langle+1|+ r_{--} |-1\rangle \langle-1|,\\
 B_2= & r_{++} |\phi^+\rangle \langle\phi^+|+ r_{+-} |\phi^-\rangle \langle\phi^-|\nonumber\\
 &+ r_{-+} |\psi^+\rangle \langle\psi^+|+ r_{--} |\psi^-\rangle \langle\psi^-|,
\end{align}
\end{subequations}
where
\begin{subequations}
\begin{align}
 &|\phi^{\pm}\rangle = \frac{1}{\sqrt{2}} (|00\rangle \pm |11\rangle),\\
 &|\psi^{\pm}\rangle = \frac{1}{\sqrt{2}} (|01\rangle \pm |10\rangle)
\end{align}
\end{subequations}
are the Bell states, which are the common eigenvectors of $\sigma _z\otimes \sigma _z$ and
$\sigma _x\otimes \sigma _x$. $A_i$ and $B_i$ have possible outcomes $r_{++}$ [representing $(+1,+1)$], $r_{+-}$, $r_{-+}$, and $r_{--}$.

In addition, Eve has to prepare 4 qubits (qubits 1 and 3 for Alice, and qubits 2 and 4 for Bob) in one of the following 16 4-qubit states: $\phi^+_{12} \phi^+_{34}$, $\phi^+_{12} \phi^-_{34}$, \ldots, $\psi^-_{12} \psi^-_{34}$. As shown in Table \ref{Table3}, each of these states violates one of the 16 Bell inequalities $\beta_{i} \stackrel{\mbox{\tiny{ LHV}}}{\leq} 7$ (for $i=1,\ldots,16$) up to its algebraic maximum, $9$, which is also the maximum value allowed by the no-signaling principle. This occurs because, from each local experiment, one of the parties can predict with certainty the other party's result. For instance, from the result of $A_0$, Alice can predict with certainty Bob's result for $B_0^{10}$, $B_1^{01}$, and $B_2^{10}$.


\begin{table*}[htb]
\caption{\label{Table3}Values of $\beta_i$ for the 16 products of Bell states. Values in bold indicate the maximum violation allowed by no-signaling.}
\begin{ruledtabular}
\begin{tabular}{r|rrrrrrrrrrrrrrrr}
 & $\beta_{1}$ & $\beta_{2}$ & $\beta_{3}$ & $\beta_{4}$ & $\beta_{5}$ & $\beta_{6}$ & $\beta_{7}$ & $\beta_{8}$
 & $\beta_{9}$ & $\beta_{10}$ & $\beta_{11}$ & $\beta_{12}$ & $\beta_{13}$ & $\beta_{14}$ & $\beta_{15}$ & $\beta_{16}$ \\
 \hline
 $\phi_{12}^+ \phi_{34}^+$ & $\bf{9}$ & $1$ & $1$ & $-3$ & $1$ & $1$ & $1$ & $-3$ & $1$ & $1$ & $1$ & $-3$ & $-3$ & $-3$ & $-3$ & $1$ \\
 $\phi_{12}^+ \phi_{34}^-$ & $1$ & $\bf{9}$ & $-3$ & $1$ & $1$ & $1$ & $-3$ & $1$ & $1$ & $1$ & $-3$ & $1$ & $-3$ & $-3$ & $1$ & $-3$ \\
 $\phi_{12}^+ \psi_{34}^+$ & $1$ & $-3$ & $\bf{9}$ & $1$ & $1$ & $-3$ & $1$ & $1$ & $1$ & $-3$ & $1$ & $1$ & $-3$ & $1$ & $-3$ & $-3$ \\
 $\phi_{12}^+ \psi_{34}^-$ & $-3$ & $1$ & $1$ & $\bf{9}$ & $-3$ & $1$ & $1$ & $1$ & $-3$ & $1$ & $1$ & $1$ & $1$ & $-3$ & $-3$ & $-3$ \\
 $\phi_{12}^- \phi_{34}^+$ & $1$ & $1$ & $1$ & $-3$ & $\bf{9}$ & $1$ & $1$ & $-3$ & $-3$ & $-3$ & $-3$ & $1$ & $1$ & $1$ & $1$ & $-3$ \\
 $\phi_{12}^- \phi_{34}^-$ & $1$ & $1$ & $-3$ & $1$ & $1$ & $\bf{9}$ & $-3$ & $1$ & $-3$ & $-3$ & $1$ & $-3$ & $1$ & $1$ & $-3$ & $1$ \\
 $\phi_{12}^- \psi_{34}^+$ & $1$ & $-3$ & $1$ & $1$ & $1$ & $-3$ & $\bf{9}$ & $1$ & $-3$ & $1$ & $-3$ & $-3$ & $1$ & $-3$ & $1$ & $1$ \\
 $\phi_{12}^- \psi_{34}^-$ & $-3$ & $1$ & $1$ & $1$ & $-3$ & $1$ & $1$ & $\bf{9}$ & $1$ & $-3$ & $-3$ & $-3$ & $-3$ & $1$ & $1$ & $1$ \\
 $\psi_{12}^+ \phi_{34}^+$ & $1$ & $1$ & $1$ & $-3$ & $-3$ & $-3$ & $-3$ & $1$ & $\bf{9}$ & $1$ & $1$ & $-3$ & $1$ & $1$ & $1$ & $-3$ \\
 $\psi_{12}^+ \phi_{34}^-$ & $1$ & $1$ & $-3$ & $1$ & $-3$ & $-3$ & $1$ & $-3$ & $1$ & $\bf{9}$ & $-3$ & $1$ & $1$ & $1$ & $-3$ & $1$ \\
 $\psi_{12}^+ \psi_{34}^+$ & $1$ & $-3$ & $1$ & $1$ & $-3$ & $1$ & $-3$ & $-3$ & $1$ & $-3$ & $\bf{9}$ & $1$ & $1$ & $-3$ & $1$ & $1$ \\
 $\psi_{12}^+ \psi_{34}^-$ & $-3$ & $1$ & $1$ & $1$ & $1$ & $-3$ & $-3$ & $-3$ & $-3$ & $1$ & $1$ & $\bf{9}$ & $-3$ & $1$ & $1$ & $1$ \\
 $\psi_{12}^- \phi_{34}^+$ & $-3$ & $-3$ & $-3$ & $1$ & $1$ & $1$ & $1$ & $-3$ & $1$ & $1$ & $1$ & $-3$ & $\bf{9}$ & $1$ & $1$ & $-3$ \\
 $\psi_{12}^- \phi_{34}^-$ & $-3$ & $-3$ & $1$ & $-3$ & $1$ & $1$ & $-3$ & $1$ & $1$ & $1$ & $-3$ & $1$ & $1$ & \bf{9} & $-3$ & $1$ \\
 $\psi_{12}^- \psi_{34}^+$ & $-3$ & $1$ & $-3$ & $-3$ & $1$ & $-3$ & $1$ & $1$ & $1$ & $-3$ & $1$ & $1$ & $1$ & $-3$ & \bf{9} & $1$ \\
 $\psi_{12}^- \psi_{34}^-$ & $1$ & $-3$ & $-3$ & $-3$ & $-3$ & $1$ & $1$ & $1$ & $-3$ & $1$ & $1$ & $1$ & $-3$ & $1$ & $1$ & \bf{9} \\
\end{tabular}
\end{ruledtabular}
\end{table*}


\section{Maximum nonlocality of systems that never interacted}


Suppose that Eve wants to prepare an equally weighted mixture of the 16 NBs. Eve can do it by producing, at will, quartets of qubits in the 16 4-qubit states with equal frequency. However, there is an alternative way to make Alice and Bob share an equally weighted mixture of NBs in which Eve's intervention is unnecessary.


\begin{figure}[t]
\vspace{-2cm}
\centerline{\includegraphics[width=0.52\textwidth]{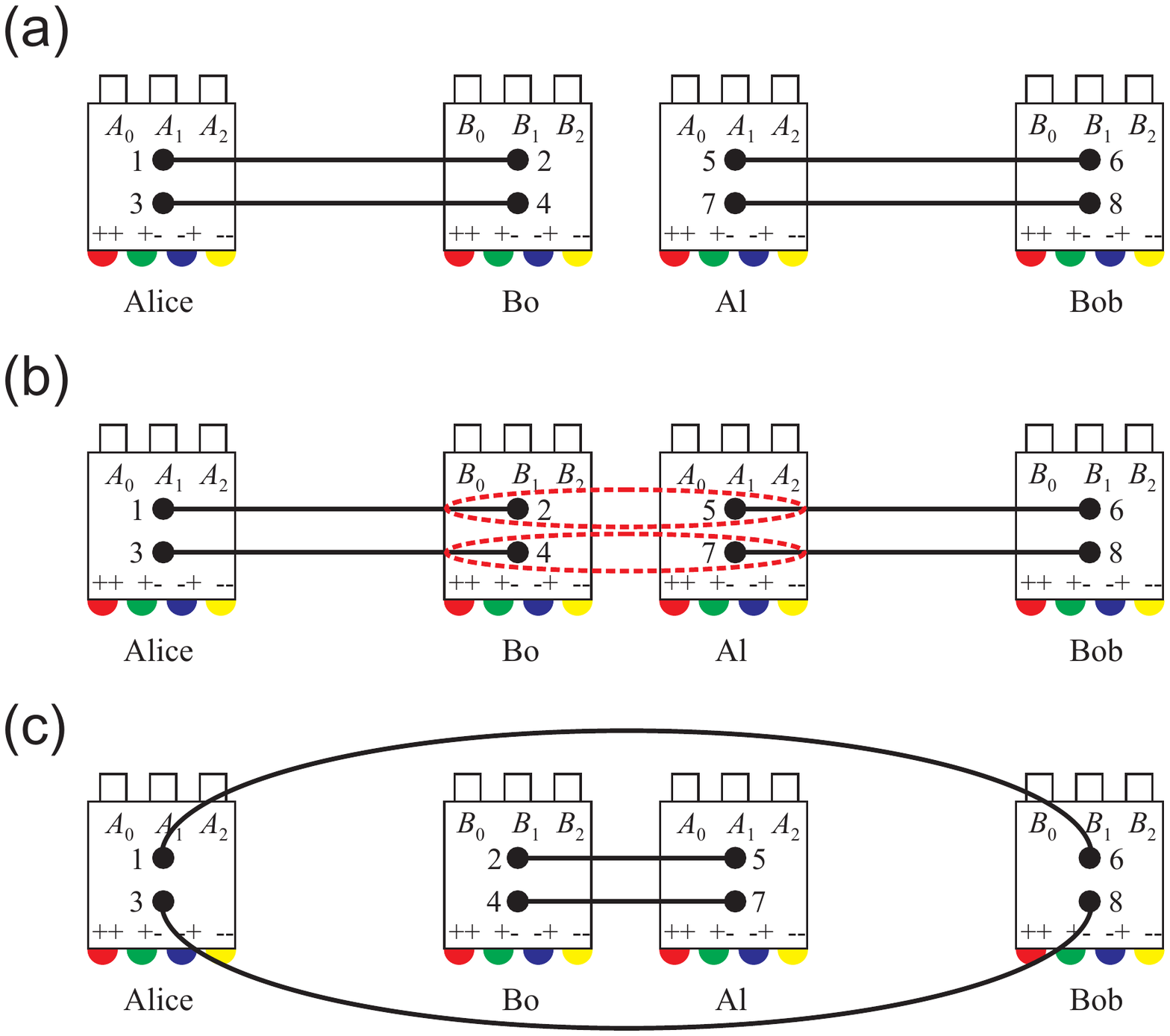}}
\caption{\label{Fig2}{\em Entanglement swapping between two pairs of 4-dimensional systems}. (a) Initially, Alice and Bo share a NB based on the observables $A_i$ and $B_j$ in (\ref{Ai}) and (\ref{Bi}), and in 4 qubits such that 1 and 2 (3 and 4) are in a Bell state. Al and Bob share a similar NB. (b) A measurement distinguishing among the 4 Bell states is performed on qubits 2 and 5, and a similar measurement is performed on qubits 4 and 7. (c) As a result, entanglement swapping occurs: Qubits 2 and 5 (4 and 7) end up in one of the 4 Bell states, and qubits 1 and 6 (3 and 8) end up in the corresponding Bell state. As a consequence, Alice and Bob end up sharing a NB without any interaction between their qubits having taken place.}
\end{figure}


This alternative is based on entanglement swapping \cite{ZZHE93} between two pairs of 4-dimensional systems; see Fig.~\ref{Fig2}. Initially, Alice and Bo share a NB based on a Bell state between qubits 1 and 2, another Bell state between qubits 3 and 4, and the observables $A_i$ and $B_j$ given by (\ref{Ai}) and (\ref{Bi}). Similarly, Al and Bob share a NB based on a Bell state between qubits 5 and 6, another Bell state between qubits 7 and 8, and the observables $A_i$ and $B_j$. Entanglement swapping occurs when a measurement distinguishing the 4 Bell states is performed on qubits 2 and 5, and a similar measurement is performed on qubits 4 and 7: Qubits 2 and 5 (4 and 7) end up in one of the 4 Bell states and qubits 1 and 6 (3 and 8) end up in the corresponding Bell state. As a consequence, Alice and Bob end up sharing a NB.


\begin{figure}[t]
 \vspace{-4cm}
 \centerline{\includegraphics[width=0.54\textwidth]{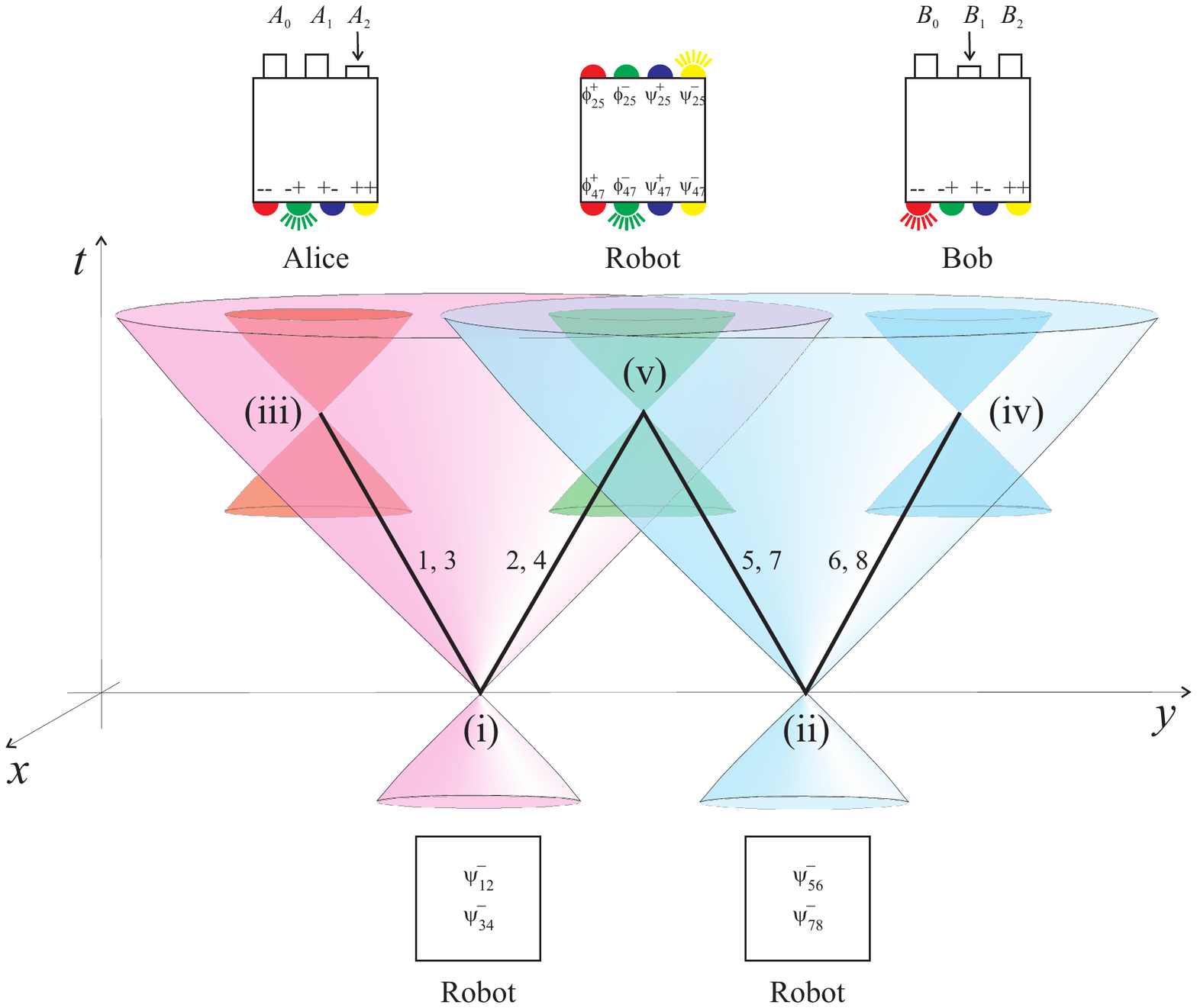}}
 \caption{\label{Fig3}{\em Maximum nonlocality of systems that never interacted}. (i) A source emits qubits 1 and 2 in the Bell state $\psi_{12}^-$ and qubits 3 and 4 in the state $\psi_{34}^-$. (ii) A similar source, synchronized with (i), emits qubits 5 and 6 in the state $\psi_{56}^-$ and qubits 7 and 8 in the state $\psi_{78}^-$. (iii) Alice performs a measurement $A_i$ on qubits 1 and 3. (iv) Bob performs a measurement $B_j$ on qubits 6 and 8. The quantum state of qubits 1, 3, 6, and 8 is maximally mixed. (v) A robot performs a measurement distinguishing the 4 Bell states on qubits 2 and 5, and a similar measurement on qubits 4 and 7, and use their results to sort the event in one of the 16 classes $\phi^+_{16} \phi^+_{38},\ldots,\psi^-_{16} \psi^-_{38}$. The region of space-time in which the robot sorts the event is spacelike separated from those in which Alice and Bob perform the local measurements.}
\end{figure}


This can be used to locally prepare equally weighted mixtures of NBs and then achieve maximum nonlocality allowed by no-signaling of systems that never interacted. The protocol is shown in Fig.~\ref{Fig3}. The process starts when two distant independent sources emit quartets of qubits in products of Bell states. In one region of space-time, Alice performs local measurements on pairs of qubits. In a causally unconnected region, Bob performs similar measurements. Notice that, so far, for any observer, the quantum state of the 4 qubits tested by Alice and Bob is maximally mixed (therefore unentangled). However, a robot, in a causally unconnected region from those in which Alice and Bob are performing the measurements, can sort the events in subsets such that each of them maximally violates one of the 16 Bell inequalities (\ref{nonlocalineq}). For that, the robot performs measurements distinguishing the 16 products of Bell states.

Contemplating the whole experiment, we notice that we ended up with a nonlocality experiment with the following properties: On one hand, the Bell test is performed on systems that never interacted. On the other hand, {\em by a measurement performed in a region spacelike separated from both Alice's and Bob's measurements}, the events are sorted in sets such that each set violates a Bell inequality up to its no-signaling limit. Therefore, the experiment reveals the maximum bipartite nonlocality allowed by no-signaling on systems that never interacted.


\section{Relation to previous results}


Entanglement swapping was used together with CHSH inequalities in the delayed entanglement swapping protocol \cite{Peres00}. However, there, Eve sorts Alice's and Bob's results in subsets not maximally violating a Bell inequality, and does it {\em after} the results are already recorded (thus Eve knows Alice's and Bob's choices).

The Bell inequality $\beta_{16} \stackrel{\mbox{\tiny{ LHV}}}{\leq} 7$ was introduced in \cite{Cabello01b} in the context of all-versus-nothing proofs. The Bell inequality $\beta_{1} \stackrel{\mbox{\tiny{ LHV}}}{\leq} 7$ has been recently used to experimentally obtain nonlocal correlations with a very small local part \cite{AGACVMC12}.


\section{Proposed experimental realization}


To experimentally prepare quantum NBs (the experiment in Fig.~\ref{Fig1}), we can use pairs of photons entangled simultaneously in two different degrees of freedom (hyperentanglement). All observables $A_i$ and $B_i$ in (\ref{Ai}) and (\ref{Bi}) have been measured using the polarization and spatial degrees of a photon on a 2-photon system hyperentangled in both polarization and spatial degrees of freedom prepared in one of the 16 states needed \cite{AGACVMC12,YZZYZZCP05}. The challenge is to achieve spatial separation between the local measurements on the two photons in order to guarantee spacelike separation (in previous experiments \cite{AGACVMC12,YZZYZZCP05} the distance between Alice and Bob is less than a meter).

The experiment to demonstrate maximum nonlocality allowed by no-signaling of systems that have not interacted (the experiment in Fig.~\ref{Fig3}) is much more demanding. It requires entanglement swapping between two pairs of 4-dimensional systems, 3 mutually spacelike separated regions (Alice's, Bob's, and the robot's), and systems in which we can measure the 4-dimensional local observables $A_i$ and $B_i$. One possibility is to achieve 4-qubit Bell state discrimination by using additional degrees of freedom, as proposed (for two-qubit Bell state discrimination) in \cite{WPM03} and demonstrated in \cite{BVMD07,WBK07}. The challenge is to extend this idea to more degrees of freedom. For example, we could achieve 4-qubit Bell state discrimination in polarization and spatial degrees of freedom by using an 8-qubit system which includes time-energy entanglement \cite{BLPK05} and also entanglement in an extra spatial degree of freedom \cite{CVDMC09}. Alternatively, other recent proposals for distinguishing the 16 states \cite{SDL10,RWHLD12} could be adopted.


\section{Conclusions}


Nature offers a much stronger form of nonlocality than the one observed in standard experiments. Here we have shown that, if the parties can choose among a larger number of measurements and outcomes, then quantum NBs providing the maximum bipartite nonlocality allowed by the no-signaling principle become possible. In addition, for these NBs, entanglement swapping allows us to prepare equally weighted mixtures of NBs between two parties that have never interacted. The final picture is that of a much stronger form of nonlocality, in which Alice and Bob perform their local measurements on systems that have never interacted, and found that every pair is as nonlocal as allowed by no-signaling. This nonlocality can be observed experimentally with current technology and may lead to new applications in quantum information processing.


\section*{Acknowledgments}


The author thanks A. Ac\'{\i}n, C. Budroni, R. Gallego, and P. Mataloni for useful conversations, C.B. and R.G. for independently checking the tightness of the Bell inequalities (\ref{nonlocalineq}) and (\ref{beta}), and A.J. L\'opez-Tarrida for his help in Figs.\ \ref{Fig1} and \ref{Fig3}. This work was supported by Project No.\ FIS2011-29400 (Spain).



\end{document}